# Martensitic relief observation by atomic force microscopy in yttria stabilized zirconia

Sylvain Deville[#], Jérôme Chevalier

INSA-GEMPPM, UMR CNRS 5510, 69621 Villeurbanne, France.

The tetragonal to monoclinic (t-m) phase transformation of zirconia has been the object of extensive investigations of the last twenty years, and is now recognised as being of martensitic nature. However, martensitic transformation has only been observed by transmission electron microscopy or indirect methods. Though the benefit on the fracture toughness and crack resistance was the main interest, the transformation is now considered for its consequences on the degradation of the material. The use of AFM reported here allowed the observation of the first stages of martensite relief growth and of new martensitic features.

## I. Introduction

The t-m phase transformation of zirconia, has been probably the most studied phase transformation among ceramics, since the transformation toughening effect increases fracture toughness and crack growth resistance[1-2]. There is now a regain of interest from a biomedical point of view, since degradation and rupture of hip joints prostheses related to the transformation have been reported recently[3]. Phase transformation occurring with time at the surface of zirconia implants is now referred as ageing. Zirconia is indeed retained in its tetragonal metastable structure at ambient temperature after processing, and phase transformation to its stable monoclinic phase might occur under hydrothermal and mechanical stresses[4-8]. Though the underlying chemical mechanism is still a point of debate, the transformation is now commonly recognised as being of martensitic nature[2,9-10], i.e. diffusionless, athermal, involving shape change dominated by shear and occurring at high speed. Indirect methods such as XRD or differential thermal analysis (DTA) provide information's on the global behaviour of the material. Further progress into the nucleation and growth (NG) mechanism understanding was performed by using optical interferometry[12] and martensitic relief observations were also reported[13] by interferometry and Nomarski interference. One AFM observation[14], performed at high scanning speed and in tapping mode has confirmed the NG nature of the transformation. Transmission electron microscopy (TEM) on thin foils has reported[1,15-16] the presence of monoclinic laths within the grains. The next step in the mechanism understanding is the observation of the transformation initiation. Thus AFM observations brought new features to light, associated with the relief, that were not accessible by other conventional techniques.

## II. Experimental procedure

Samples were processed using yttria-stabilized zirconia powder (3Y-TZP, Tosoh TZ-3YS, Tosoh corporation, Tokyo, Japan). Green bodies, after uniaxial cold pressing at 10 bars, were isostatically cold pressed at 1300 bars and finally sintered at 1773 K for five hours in air. The obtained plates were machined to small bars. The side of each bar on which analysis was to be performed was mirror-polished with diamond slurries and pastes, reaching surface roughness (Ra) values as low as 2 nm. Samples were thermally etched for 18 min at 1673 K, i.e. 100 K below the sintering temperature, to form slight thermal graves of grain boundaries without affecting the physical properties. The ageing treatments were conducted in an autoclave (Fisher Bioblock, France) at 383 K, ensuring a 100% water vapour atmosphere. The surface was observed using atomic force microscopy (D3100, Digital Instruments) in contact mode with a scanning speed of 10 µm s$^{-1}$.

---

[#]Corresponding author : sylvain.deville@insa-lyon.fr

**III.    Results and discussion**

A limited transformed zone is shown on **Fig. 1**. The AFM deflection image provides a very clear image of the surface, showing typical martensitic self accommodating variants within almost each grain of the zone, after 1 hour of ageing treatment. Most of the theories[10] developed so far predict the apparition of a regular pattern of martensitic laths within a grain, related to the crystallographic orientation of the grain in a very straightforward manner. The observations here clearly show the apparition of different orientations of the martensitic relief for each grain. Image of **Fig. 2** was acquired on a single grain and at very low scanning speed to provide a maximum definition of the picture. The very regular orientation of the planes is readily observed, and different variants of various orientations and sizes are present. Though the grain is not very large, several laths within just a grain appeared during the ageing treatment. Moreover, it is worth noticing that the three main variants pairs are intersecting at a right angle. This observation is consistent the crystallography, since two perpendicular planes of the tetragonal system are crystallographically equivalent.

Another point of debate was whether a grain would transform just at once, due to high-speed transformation, or if the martensitic plates would progressively invade the grain. Some TEM observations[1,17] of partially transformed grain have already been reported. However, due to the intrinsic nature of TEM, the observed samples might be considered as being in a modified physical and chemical environment. Grains lying along the border do not have neighbouring grains neither on their side, nor above or below them. Thus the extrapolation to massive samples might be questionable[10]. This phenomenon is reported here for the first time on the surface, by means of direct observations, as shown on **Fig. 3**, where the first stage of growth of some monoclinic plates is observed firstly at the grain boundary and propagate later on to the adjacent part of the grain. This clearly proves that the transformation of a single grain does not occur instantaneously, completing previous TEM observations. It is was also expected from calculations and stress analysis[18-19] that the transformation starts at a triple junction of grain boundaries, due to higher residual stresses, though no observations have been reported so far. This is observed on **Fig. 3**. Further statistical analysis will be performed to confirm this.

Finally, several theoretical studies have shown the propagation mechanism of martensitic plates from one grain to another one, considering that high local stresses appears at the junction of the martensitic laths due to shear strain accommodation. Thus, the apparition of a single martensitic variant within a grain might provoke the transformation of another variant of the neighbouring grain, providing enough stresses-related energy to overcome the nucleation barrier. Considering this mechanism, the orientation of most of the transformed laths is different of that of the initial one. However, an interesting feature is characterised here, on **Fig. 4**. A transformed lath starts at one grain and goes through the other side of the grain boundary. Theoretically, it has been shown [10,13] that under certain crystallographic conditions of the original t phase, the transformed martensitic variant might traverse several tetragonal domains. This would explain very nicely the observed feature.

IV.    Conclusion

Observations of martensitic relief of a few dozens of nanometers in yttria stabilized zirconia by atomic force microscopy in contact mode are reported here for the first time. The different features showed here fit very well the model of martensitic transformation. The junction planes angle correspondence and the progressive invasion of martensite laths with different orientation was observed. Transgranular monoclinic variants also appeared, as predicted theoretically by some authors. Using AFM should bring observations of martensitic features that were not accessible by other techniques.

**References**

1. D. J. Green, R. H. J. Hannink, M. V. Swain *in Transformation toughening of ceramics* (CRC Press, Boca Raton, FL, 1989)
2. A. G. Evans, A. H. Heuer, Review - Transformation toughening in ceramics: martensitic transformation in crack-tip stress fields, *J. Am. Ceram. Soc*., **63(5-6)** 241-248 (1980)
3. www.ceramic-artificial-hip-implant-recall.com
4. S. Lawson, Environmental degradation of zirconia ceramics, *J. Eur. Ceram. Soc,* **13 (6)** 485-502 (1995)
5. H. Tsubakino, M. Hamamoto, R. Nozato, Tetragonal to monoclinic phase transformation during thermal cycling and isothermal ageing in yttria-partially stabilized zirconia, *J. Mater. Sci*., **26** 5521-26 (1991)
6. F. F. Lange, G.L. Dunlop, B.I. Davis, Degradation during ageing of transformation toughened $ZrO_2$-$Y_2O_3$ materials at 250°C, *J. Am. Ceram. Soc*., **69** 237-240 (1986)
7. T. Sato, M. Shimada, Transformation of yttria-doped tertragonal $ZrO_2$ polycrystals by annealing in water, *J. Am. Ceram. Soc*., **68(6)** 356-59 (1985)
8. J. Li, L. Zhang, Q. Shen, T. Hashida, Degradation of yttria stabilised zirconia at 370 K under a low applied stress, *Mater. Sci. Eng*., **A297** 26-30 (2001)
9. R. Guo, D. Guo, D. Zhao, Z. Yang, Y. Chen, Low temperature ageing in water vapor and mechanical properties of ZTA ceramics, *Mater. Lett*., **86(1)** 200-202 (2003)
10. P. M. Kelly, L. R. Francis Rose, The martensitic transformation in ceramics - its role in transformation toughening, *Prog. Mater. Sci*., **47** 463-557 (2002)
11. F. F. Lange, Transformation toughening, Part 1: Size effects associated with the thermodynamics of constrained transformations, *J. Mater. Sci*., **17** 255-262 (1982)
12. J. Chevalier, B. Calès, J. M. Drouin, Low temperature aging of Y-TZP ceramics, *J. Am. Ceram. Soc*., **82(8)** 2150-54 (1999)
13. M. Hayakawa, K. Adachi, M. Oka, Crystallographic analysis of the monoclinic herringbone structure in an arc-melted $ZrO_2$-2mol%$Y_2O_3$ alloy, *Acta Met. Mater*., **38(9)** 1753-1759 (1990)
14. H. Tsubakino, Y. Kuroda, M. Niibe, Surface relief associated with isothermal martensite in zirconia 3 mol% yttria ceramics observed by atomic force microscopy, *Com. Am. Ceram. Soc*., **82(10)** 2921-23 (1999)
15. I. W. Chen, Y. H. Chiao, Martensitic nucleation in $ZrO_2$, *Act. Metall*., **31(10)** 1627-1638 (1983)
16. Y. H. Chiao, I. W. Chen, Martensitic growth in $ZrO_2$ - an in situ, small particles, TEM study of a single interface transformation, *Act. Metall. Mater*., **38(6)** 1163-74 (1990)
17. E. P. Butler, A. H. Heuer, X-Ray microanalysis of $ZrO_2$ particles in $ZrO_2$-toughened $Al_2O_3$, *Com. Am. Ceram. Soc*., **65(12)** C206-C207 (1982)

18. L. Gremillard, T. Epicier, J. Chevalier, G.Fantozzi, Microstructural study of silica-doped ceramics, *Acta Mater.*, **48** 4647-4652 (2000)
19. S. Schmauder, H. Schubert, Significance of internal stresses for the martensitic transformation in yttria-stabilized tetragonal zirconia polycrystals during degradation, *J. Am. Ceram. Soc.*, **69(7)** 534 (1986)

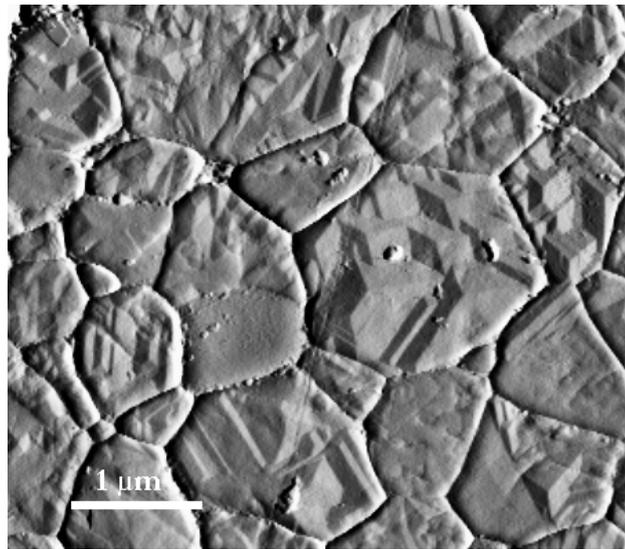

Fig. 1 : AFM deflection image. The observed relief is undoubtedly characteristic of a martensitic transformation.

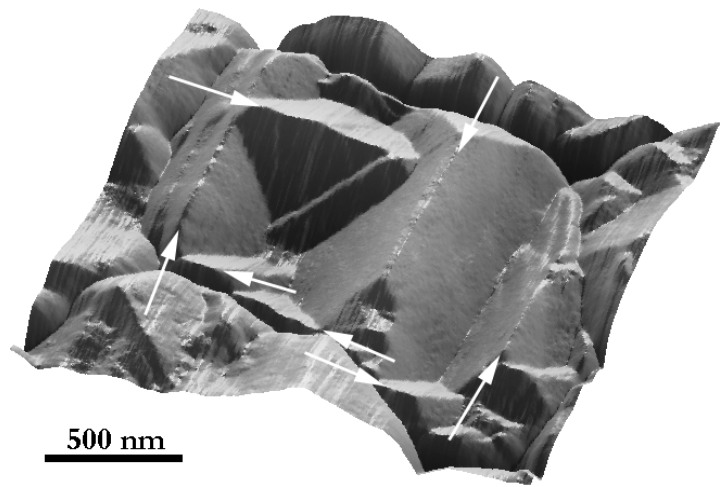

Fig. 2: AFM height image of a grain (vertical scale 80 nm). Martensitic variants crossing at right angle are observed. Arrows indicate the junction plane orientation.

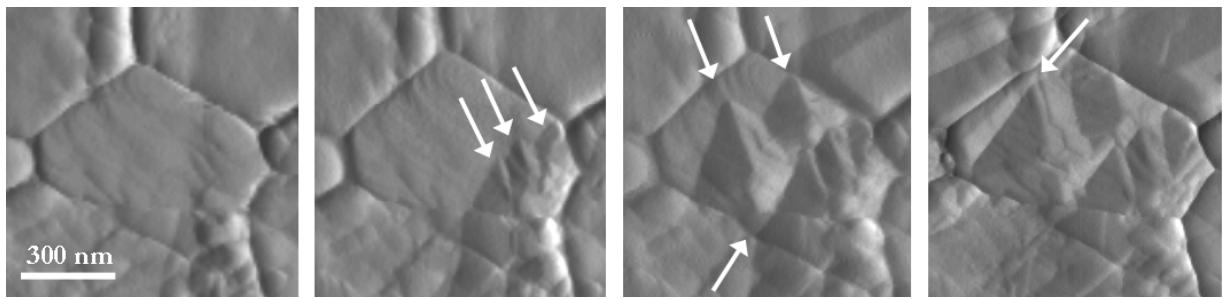

Fig. 3 : AFM deflection images at two different ageing time (20, 25, 30 and 40 min at 383 K). The growth and progressive invasion (arrows) of martensitic variants within a single grain is clear.

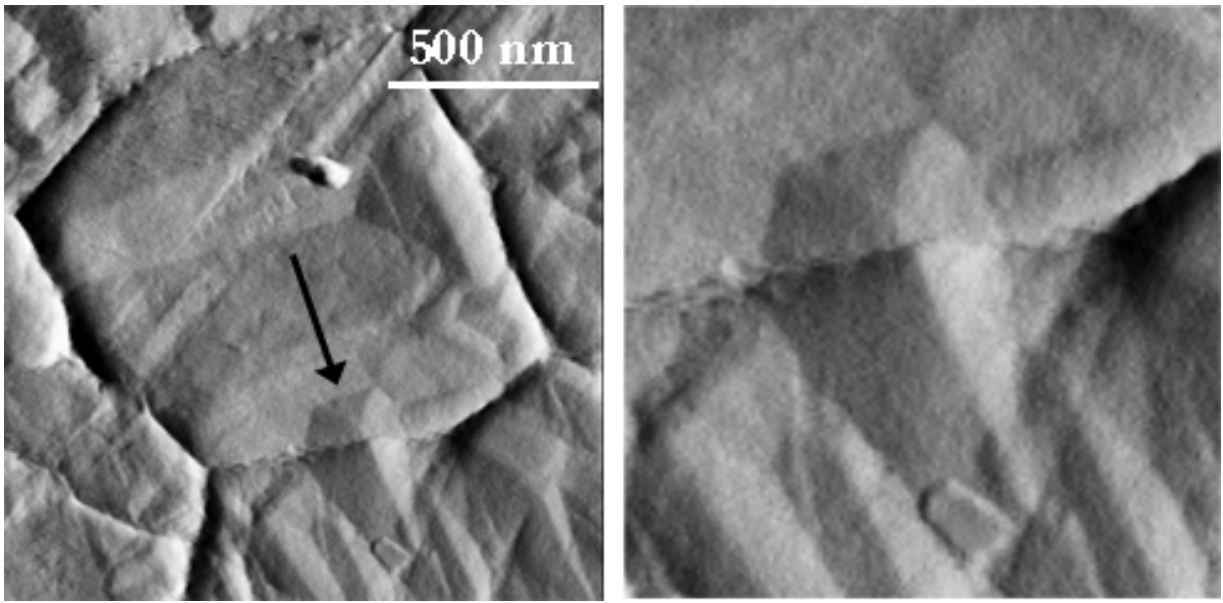

Fig. 4 : AFM deflection image. A detailed zone is extracted on the right, showing a transgranular martensitic variant.